\newcommand{\be}{\begin{equation}}
\newcommand{\ee}{\end{equation}}
\newcommand{\bea}{\begin{eqnarray}}
\newcommand{\eea}{\end{eqnarray}}
\newcommand{\nn}{\nonumber}
\renewcommand{\l}{\left}
\newcommand{\r}{\right}
\newcommand{\half}{\frac{1}{2}}
\newcommand{\inv}[1]{\frac{1}{#1}}
\newcommand{\lsim}{\mathrel{\rlap{\lower4pt\hbox{\hskip1pt$\sim$}}
                   \raise1pt\hbox{$<$}}}
\newcommand{\gsim}{\mathrel{\rlap{\lower4pt\hbox{\hskip1pt$\sim$}}
                   \raise1pt\hbox{$>$}}}
\newcommand{\hz}{H_0}
\newcommand{\psip}{\psi_+}
\newcommand{\psim}{\psi_-}
\renewcommand{\epsilon}{\varepsilon}
\documentstyle[12pt]{article}
\begin{document}
\title{General Solutions for Tunneling of \\ Scalar Fields
       with Quartic Potentials \\ in de Sitter Space}
\author{Misao Sasaki, Ewan D. Stewart and Takahiro Tanaka \\
        Department of Physics \\ Kyoto University \\
        Kyoto 606, Japan}
\maketitle
\begin{abstract}
The tunneling rates for scalar fields with quartic potentials in
de Sitter space in the limit of no gravitational back reaction are
calculated numerically and the results are fitted by analytic
formulae.
\end{abstract}
\vspace*{-77ex}
\hspace*{\fill}{\bf KUNS 1245}\hspace*{2.9em}\\
\hspace*{\fill}{To be published in}\hspace*{0.7em}\\
\hspace*{\fill}{Physical Review D}
\thispagestyle{empty}
\setcounter{page}{0}
\newpage
\setcounter{page}{1}

\section{Introduction}

The decay of a false vacuum and/or creation of topological defects
through quantum tunneling in the early Universe may play an
important role in determining the subsequent structure of the
Universe. These phenomena are particularly important in the context
of inflationary cosmology.
The nucleation of true vacuum bubbles associated with
false vacuum decay is incorporated in several viable models of
inflation as a mechanism to terminate the inflationary expansion of
the Universe \cite{Inflation}. Quantities such as the efficiency of
thermalization by bubble collisions or the degree of spatial
inhomegeneities crucially depend on the nucleation rate. Also,
topological defects such as domain walls, strings, or monopoles may
be created during the inflationary stage of the Universe \cite{Basu}.
In order to quantify their effects on the structure of the Universe,
it is necessary to know their formation rates.

The study of false vacuum decay was initiated by
Voloshin, Kobzarev, and Okun \cite{Voloshin},
and subsequently developed by Coleman \cite{Coleman},
who showed how to calculate the tunneling rate by applying the
instanton method to a scalar field having two nondegenerate minima.
According to this method, the nucleation rate is crucially determined
by the value of the action of a nontrivial classical Euclidean
solution having O(4) symmetry, called the bounce solution.
Coleman also argued that the formation and evolution of
the true vacuum bubble are described by the analytic continuation of
the bounce solution to Lorentzian time.
Then Coleman and De Luccia \cite{Luccia} made an attempt to include
gravitational effects in the bounce and obtained several important new
results.
However, their explicit results assume the thin-wall approximation,
under which the space outside the bubble wall is exactly in the false
vacuum. Hence it was unnecessary to know the behavior of the bounce
solution there.

Once the thin-wall approximation is relaxed, it becomes indispensable
to consider the global structure of the bounce solution.
In particular, when the Euclidean Einstein equations are solved
simultaneously, a finite vacuum energy density in the false vacuum
inevitably gives the topological structure of $S^4$ for the bounce
solution, leading to the absence of the asymptotic infinity in the
Euclidean space.
Subsequent studies by various people have shown that this fact
can cause very different behavior of the bounce solution
compared with the case of a flat Euclidean background.
For example, in the other extreme limit in which the width of
the bubble wall would be very large, the solution reduces to
a perfectly homogeneous $S^4$ with the value of the scalar field
staying at the top of the potential barrier. This is known as the
Hawking-Moss solution \cite{Hawking}.
Another interesting case is that of a potential with degenerate
minima.
A bounce solution with finite action exists in this case because of
the compactness of the four-volume, contrary to the case of a flat
Euclidean background.
This describes the formation of a topological defect (a domain wall
in the case of a single real scalar field) \cite{Basu}.

In the thin-wall and Hawking-Moss limits one can evaluate the bounce
action analytically.
However, for intermediate cases it must be evaluated numerically.
Thus, considering the important role played by the bounce action,
it is desirable and useful to have a formula to evaluate the general
bounce action ready to our hands.

In this paper, we consider a real scalar field with a general quartic
potential, and evaluate numerically the bounce action by varying the
parameters. Then, using the results, we construct fitting formulae
which can be applied to wide ranges of the parameter space.
This is a generalisation of the procedure of Ref.~\cite{Adams}
which was only valid in the flat space limit.
In Section 2 we introduce the formalism and derive our basic
equations. In Section 3 we derive analytic results for the Hawking-Moss
and thin-wall limits. In Section 4 we present our numerical results and
various analytic fitting formulae.

\section{Method}
\subsection{General tunneling formalism}

Following the presciption given by Coleman and De Luccia \cite{Luccia},
the tunneling rate in de Sitter space in the semiclassical limit is
given by
\be
\Gamma \propto e^{-B/\hbar} \,,
\ee
where $B$ is the bounce action. It is defined to be
\be
B = S_{\rm E} [ \phi_{\rm I} ({\bf x},\tau) ] - S_{\rm E} [\phi_{0}] \,,
\ee
where $S_{\rm E}$ is the Euclidean action, $\tau=it$ is the Euclidean time,
$\phi_{0}$ is the false vacuum value of the scalar field,
and $\phi_{\rm I}({\bf x},\tau)$ is the instanton \cite{AOS} with minimum
Euclidean action.
An instanton is a solution of the Euclidean equations of motion
which interpolates from one side of the potential barrier to the
other.\footnote{The Hawking-Moss instanton is the degenerate case of
this in that it sits on top of the potential barrier.} The instanton
with minimum Euclidean action will dominate the tunneling rate. It
corresponds to the lowest saddle point in the Euclidean action, whose
single negative mode interpolates between the false and true vacuum
states. In flat space it can be proved that the instanton has O(4)
symmetry \cite{O4} and it is usually assumed that this is also the case
in de Sitter space \cite{Luccia}. The Euclidean action for a scalar
field minimally coupled to gravity is
\be
S_{\rm E} = \int d^{4}x\, \sqrt{g} \l[ -\inv{16\pi G} R
+ \half g^{\mu\nu} \partial_{\mu} \phi \partial_{\nu} \phi + V(\phi)
\r] \,.
\ee
Assuming O(4) symmetry gives
\be
ds^2 = d\xi^2 + a^2 (\xi) \,d\Omega^2 \,,\;\;\;\;
\phi = \phi(\xi) \,,
\ee
and
\be
S_{\rm E} = 2\pi^{2} \int d\xi
\l\{ -\frac{3}{8\pi G} a \l( \dot{a}^{2} + 1 \r)
+ a^{3} \l[ \half \dot{\phi}^{2} + V(\phi) \r] \r\} \,.
\ee

\subsection{A quartic potential}
In this paper we limit ourselves to the case of a quartic potential
with a local minimum (the false vacuum) at $\phi=0$, a local maximum
(the top of the potential barrier) at $\phi=\phi_{+}$, and a global
minimum (the true vacuum) at $\phi=\phi_{-}$. The potential can
be written as
\be
V(\phi)=V_{0}+\half m^{2}\phi^{2}-\mu\phi^{3}+\lambda\phi^{4} \,.
\ee
Using the following change of variables we can eliminate two of the
five parameters from the dynamics:
\be
\psi=\frac{\mu}{m^{2}}\phi \,,\;\;\;\; \rho=ma
\,,\;\;\;\; \zeta=m\xi \,,
\ee
\be
\hz=\inv{m}\sqrt{\frac{8\pi GV_{0}}{3}} \,,\;\;\;\;
\nu=\frac{2\lambda m^{2}}{\mu^{2}} \,,
\ee
and
\be
\kappa = \frac{8\pi Gm^{4}}{\mu^{2}}=\frac{8\pi}{m_{\rm Pl}^{2}}
\frac{m^{4}}{\mu^{2}} \,.
\ee
It will also be useful to introduce the notation
\be
\label{eps}
\epsilon = 1 - \nu \,.
\ee
Then
\be
\label{SE3}
S_{\rm E} = \frac{2\pi^{2}m^{2}}{\mu^{2}}\int d\zeta
\l\{-\frac{3}{\kappa}\rho\l(\dot{\rho}^{2}+1-\hz^{2}\rho^{2}\r)
+ \rho^{3}\l[\half\dot{\psi}^{2}+U(\psi)\r]\r\} \,,
\ee
where
\be
\label{U}
U(\psi) = \half \psi^{2} - \psi^{3} + \half \nu \psi^{4} \,.
\ee
The Euclidean equations of motion are
\be
\ddot{\psi}+3\frac{\dot{\rho}}{\rho}\dot{\psi}=U' \,,
\ee
\be
\dot{\rho}^{2}-1+\hz^{2}\rho^{2}=\frac{\kappa}{3}\rho^{2}
\l(\half\dot{\psi}^{2}-U\r) \,,
\ee
and the bounce action is
\be
B = S_{\rm E}\l[\psi(\zeta),\rho(\zeta)\r]
-S_{\rm E}\l[0,\inv{\hz}\sin\hz\zeta\r] \,.
\ee
Note that $U(\psi)$ has its local maximum and global minimum at
\be
\psi_{\pm}=\frac{3}{4\nu}\l(1\mp\sqrt{1-\frac{8}{9}\nu}\r) \,,
\ee
respectively, and $U(\psim)=0$ when $\nu=1$ so that $0\leq\nu\leq1$.

\subsection{Gravitational back reaction}
The gravitational back reaction will be negligible if $\kappa\ll 1$
except for in the thin-wall limit, $\epsilon \ll 1$ and
$\hz \ll 1$ (see Section \ref{tw}). The bounce action for a general
potential in the thin-wall limit has been given in Ref.~\cite{Parke},
and in our case it reduces to
\be
B = \frac{m^2}{\mu^2} \frac{\pi^{2}}{3}
\inv{ \sqrt{ \l(\epsilon-\inv{24}\kappa\r)^{2} + \hz^{2} } \l[
\l( \epsilon + \sqrt{\l(\epsilon-\inv{24}\kappa\r)^{2}+\hz^{2}} \r)^2
- \l( \inv{24} \kappa \r)^2 \r] } \,.
\ee
We see that the gravitational back reaction is negligible for
$\kappa\ll\hz$ or $\kappa\ll\epsilon$.
Thus for $\kappa \ll \min\{1,\max\{\hz,\epsilon\}\}$ the gravitational
back reaction is negligible and we will assume this throughout the rest
of the paper. The Euclidean equations of motion then become
\be
\rho=\inv{\hz}\sin\hz\zeta \,,
\ee
and
\be
\label{eem}
\ddot{\psi}+3\hz\cot\hz\zeta\,\dot{\psi}=U' \,,
\ee
and the bounce action becomes
\be
\label{ba}
B = \frac{m^2}{\mu^2} \frac{2\pi^2}{\hz^{3}}
\int_{0}^{\frac{\pi}{\hz}}d\zeta\,\sin^{3}\hz\zeta
\l[\half\dot{\psi}^{2}+U(\psi)\r] \,.
\ee
The dynamics now depends on only two parameters, $\nu$ and $\hz$, and
so a numerical treatment becomes tractable. See Figure 1 for a
description of the parameter space and a sketch of our results.

\section{Analytic Results}
\subsection{The Hawking-Moss instanton}
\label{hm}
The Euclidean equation of motion, Eq.~(\ref{eem}), always has the trivial
solution $\psi=\psip$. This is called the Hawking-Moss instanton.
In flat space it has infinite bounce action and plays no role, but
because Euclidean de Sitter space has a finite volume,
$8\pi^{2}/3H^{4}$, it has a finite bounce action
\be
\label{hmb}
B = \frac{m^2}{\mu^2} \frac{8\pi^2}{3\hz^{4}} U(\psip)
\ee
in de Sitter space and is the dominant instanton for sufficiently
large $\hz$.
For our simple case of a quartic potential the condition under which
the Hawking-Moss instanton dominates can be obtained by the perturbative
analysis below \cite{Jensen}, but for a general potential this will
not be valid because the bounce action depends on the global properties
of the potential and not just on the local properties around the top of
the potential barrier.

Let $\omega^{2} \equiv -U''(\psip) = 2-3\psip$ and $\chi=\psi-\psip$.
Assume $\chi\ll 1$.
Then, from Eq.~(\ref{eem}),
\be
\ddot{\chi}+3\hz\cot\hz\zeta\,\dot{\chi}+\omega^{2}\chi \simeq 0 \,.
\ee
This equation has a nontrivial regular solution only for $\omega^{2}
=n(n+3)\hz^{2}$ where $n=0,1,2,\ldots$ . For $\omega=0$ it is
$\chi= {\rm constant}$. For $\omega \ll \hz$, we expect it still to be
an approximate solution, in which case $B$ is maximised for $\chi=0$.
Note that we must maximise $B$ to obtain a solution that interpolates
from one side of the potential barrier to the other. For $\omega
=2\hz$ the regular solution is
\be
\chi = A\cos\hz\zeta \,,
\ee
and for $\omega \simeq 2\hz$ we expect it still to be an approximate
solution.
Then for $A$ sufficiently small, $A^{2} \ll \omega^{2}-4\hz^{2}$,
\be
B = \frac{m^2}{\mu^2} \frac{8\pi^{2}}{3\hz^{4}} \l\{ U(\psip)
- \inv{10} \l(\omega^{2}-4\hz^{2}\r) A^{2} \r\} \,.
\ee
Therefore for $\omega^{2}<4\hz^{2}$, $B$ is minimised for $A=0$,
{\mbox i.e.}, the Hawking-Moss instanton; while for
$\omega^{2}>4\hz^{2}$, $B$ is minimised for $A\rightarrow\infty$,
{\mbox i.e.}, the Hawking-Moss instanton is no longer the dominant
instanton. Thus the Hawking-Moss instanton dominates for
\be
\label{hmbdy}
\hz^2 \geq \inv{2} \l( 1 - \frac{3}{2} \psip \r) \,.
\ee

Now we will calculate the first-order corrections to Eq.~(\ref{hmb})
for $0 < \omega^{2}-4\hz^{2} \ll 1$. From Eqs.~(\ref{U}) and
(\ref{eem}),
\be
\ddot{\chi}+3\hz\cot\hz\zeta\dot{\chi}+4\hz^{2}\chi =
-\l(\omega^{2}-4\hz^{2}\r)\chi
-3\l(1-2\nu\psip\r)\chi^{2}+2\nu\chi^{3} \,.
\ee
This has the approximate solution \cite{Tanaka}
\be
\chi = A\cos\hz\zeta + \frac{1-2\nu\psip}{2-3\psip}A^{2}\cos2\hz\zeta
+ O\l(A^{3}\cos3\hz\zeta\r) \,,
\ee
where
\be
A^{2} = \frac{7\l(2-3\psip\r)}{6\l(1-\nu\psip\r)}
\l(\omega^{2}-4\hz^{2}\r)+O\l(\l(\omega^{2}-4\hz^{2}\r)^{2}\r) \,.
\ee
Substituting into Eq.~(\ref{ba}) then gives
\be
B = \frac{m^2}{\mu^2} \frac{8\pi^{2}}{3\hz^{4}} \l[ U(\psip)
- \frac{7\l(2-3\psip\r)}{120\l(1-\nu\psip\r)}
\l(\omega^{2}-4\hz^{2}\r)^{2} + \ldots \r] \,.
\ee
For $\nu=0$ this gives
\be
\label{Bnz}
B = \frac{m^2}{\mu^2} \frac{4\pi^{2}}{81\hz^{4}}
\l[ 1 - \frac{63}{20}\l(1-4\hz^{2}\r)^{2} + \ldots \r] \,,
\ee
and for $\nu=1$ it gives
\be
\label{Bno}
B = \frac{m^2}{\mu^2} \frac{\pi^{2}}{12\hz^{4}}
\l[ 1 - \frac{7}{15}\l(1-8\hz^{2}\r)^{2} + \ldots \r] \,.
\ee

\subsection{The thin-wall approximation}
\label{tw}
{}From Eqs.~(\ref{eps}) and (\ref{U}),
\be
U(\psi) = \half\psi^{2}(1-\psi)^{2} - \half\epsilon\psi^{4} \,.
\ee
In the thin-wall approximation, $\epsilon \ll 1$ so that the
true and false vacua are nearly degenerate, and $\hz \ll 1$ so
that a large bubble can fit into the Euclidean de Sitter space.
The instanton can then be divided into three regions
\begin{enumerate}

\item
Inside the bubble where
\be
\psi = \psim \simeq 1 \;\;\;\;{\rm and}\;\;\;\;
U\simeq-\half\epsilon \,.
\ee

\item
The bubble wall where
\be
\zeta \simeq \zeta_{\rm W} \,,\;\;\;\; \ddot{\psi} \simeq U'
\;\;\;{\rm and}\;\;\; U \simeq \half \psi^{2}(1-\psi)^{2} \,.
\ee
Therefore
\be
\label{wall}
\dot{\psi} \simeq -\sqrt{2U} \simeq -\psi(1-\psi) \;\;\;\;\;\;\l(
{\rm therefore} \;\;\; \psi=\inv{1+e^{\zeta-\zeta_{\rm W}}} \r) \,.
\ee

\item
Outside the bubble where
\be
\psi=0 \,.
\ee

\end{enumerate}
Putting it all into Eq.~(\ref{ba}) then gives
\bea
B & = & \frac{m^2}{\mu^2} \frac{2\pi^{2}}{\hz^{3}} \l[ -\half\epsilon
\int_{0}^{\zeta_{\rm W}}d\zeta\,\sin^{3}\hz\zeta
+\sin^{3}\hz\zeta_{\rm W}\int_{0}^{1}d\psi\,\psi(1-\psi)\r] \,, \\
  & = & \frac{m^2}{\mu^2} \frac{2\pi^{2}}{\hz^{3}}
\l[ - \half \frac{\epsilon}{\hz}
\l( \frac{2}{3} - \cos\hz\zeta_{\rm W}
	+ \frac{1}{3}\cos^{3}\hz\zeta_{\rm W} \r)
+ \inv{6} \sin^{3} \hz \zeta_{\rm W} \r] \,,
\eea
and maximising $B$ with respect to $\zeta_{\rm W}$ gives
\be
\label{wr}
\tan\hz\zeta_{\rm W} = \frac{\hz}{\epsilon} \,.
\ee
Note that $\rho_{\rm W}=(1/\hz)\sin\hz\zeta_{\rm W}
=1/\sqrt{\epsilon^{2}+\hz^{2}}\gg 1$.
Substituting back into $B$ gives
\be
\label{twb0}
B = \frac{m^2}{\mu^2} \frac{\pi^2}{3} \inv{ \sqrt{\epsilon^2+\hz^{2}}
\l( \epsilon + \sqrt{\epsilon^2+\hz^{2}} \r)^2 } \,.
\ee

Now we will calculate the first-order corrections in $\epsilon$ and
$\hz$ to Eq.~(\ref{twb0}). As before, divide the instanton into three
parts.
\begin{enumerate}

\item
Inside the bubble where
\be
\psi = \psim \simeq 1+2\epsilon \;\;\;{\rm and}\;\;\;
U \simeq -\half\epsilon(1+4\epsilon) \,.
\ee

\item
The bubble wall where, from Eqs.~(\ref{eem}), (\ref{wall}), and
(\ref{wr}),
\be
\ddot{\psi} \simeq U'-3\epsilon\dot{\psi}
\simeq U'+3\epsilon\psi(1-\psi) \,.
\ee
Therefore
\be
\label{pde}
\dot{\psi} \simeq -(1-\half\epsilon)\psi(1-\psi) - 2\epsilon\psi \,,
\ee
and
\be
\half\dot{\psi}^{2} + U \simeq (1-\epsilon)\psi^{2}(1-\psi)^{2}
+ \half\epsilon\psi^{2}(5-6\psi) \,.
\ee
Therefore $\dot{\psi}^{2}/2 + U = O(\epsilon^{2})$ when
$\psi = O(\epsilon)$ and
$\dot{\psi}^{2}/2 + U = -\epsilon/2 + O(\epsilon^{2})$
when $\psi = 1 + O(\epsilon)$.
Therefore from Eq.~(\ref{wall}) the boundaries of the bubble wall
can be taken to be at $\zeta=\zeta_{\rm W}\pm\ln\epsilon$.

\item
Outside the bubble where
\be
\psi=0 \,.
\ee

\end{enumerate}
Putting it all into Eq.~(\ref{ba}) then gives
\bea
B & = & \frac{m^2}{\mu^2} \frac{2\pi^{2}}{\hz^{3}}
\l\{ - \half \epsilon (1+4\epsilon)
\int_{0}^{\zeta_{\rm W}+\ln\epsilon} d\zeta\, \sin^{3} \hz \zeta
\r. \nonumber \\
 & & \l.+\int_{\zeta_{\rm W}+\ln\epsilon}^{\zeta_{\rm W}
-\ln\epsilon}d\zeta\,\sin^{3}\hz\zeta\l[(1-\epsilon)\psi^{2}
(1-\psi)^{2}+\half\epsilon\psi^{2}(5-6\psi)\r]\r\} \,,
\eea
which from Eqs.~(\ref{wall}) and (\ref{pde}), and after noting that we can
use Eq.~(\ref{wr}) because $\partial B / \partial \zeta_{\rm W} = 0$,
eventually gives
\be
\label{twb1}
B= \frac{m^2}{\mu^2} \frac{\pi^{2}}{3} \inv{ \sqrt{\epsilon^2+\hz^{2}}
\l(\epsilon+\sqrt{\epsilon^{2}+\hz^{2}}\r)^{2}}
\l[ 1 + \frac{11}{2}\epsilon
+ \frac{3\epsilon^2}{\sqrt{\epsilon^{2}+\hz^{2}}}
+ \frac{3\epsilon^3}{2\l(\epsilon^{2}+\hz^{2}\r)} \r] \,.
\ee
In the limit, $\hz\ll\epsilon$, this gives
\be
B = \frac{m^2}{\mu^2} \frac{\pi^2}{12\epsilon^3}
\l(1+10\epsilon-\frac{\hz^{2}}{\epsilon^{2}}+\ldots\r) \,,
\ee
while in the limit $\epsilon\ll\hz$ it gives
\be
B = \frac{m^2}{\mu^2} \frac{\pi^2}{3\hz^{3}}
\l( 1 - 2\frac{\epsilon}{\hz} + \ldots \r) \,,
\ee
so that we need to calculate to a higher order in $\hz$ in this
limit to get the first-order correction. We will now do this.

Set $\epsilon=0$. Then
\be
U=\half\psi^{2}(1-\psi)^{2} \,,
\ee
and by symmetry $\zeta_{\rm W} = \pi / 2\hz$.
Therefore, from Eq.~(\ref{eem}),
\bea
\dot{\psi}^{2} & = & 2U-6\hz\int d\zeta\,\cot\hz\zeta\,\dot{\psi}^{2}
\,,  \nonumber \\
 & = & 2U\l[1+O\l(\hz^{2}\r)\r] \,.
\eea
Therefore
\be
\frac{\half\dot{\psi}^{2}+U}{\dot{\psi}}
=-\sqrt{2U}\l[1+O\l(\hz^{4}\r)\r]  \,.
\ee
Therefore from Eqs.~(\ref{ba}) and (\ref{wall}),
\bea
B & = & \frac{m^2}{\mu^2} \frac{2\pi^2}{\hz^{3}} \int_{0}^{1} d\psi\,
\l[1-\frac{3}{2}\hz^{2}\ln^{2}\l(\inv{\psi}-1\r)\r]\psi(1-\psi)
+ \ldots \,, \nonumber \\
\label{twe}
  & = & \frac{m^2}{\mu^2} \frac{\pi^2}{3\hz^{3}}
\l[ 1 - \half\l(\pi^2-6\r)\hz^{2} + \ldots \r] \,.
\eea

\section{Numerical Results}

We solved Eq.~(\ref{eem}) numerically with $\nu$ and $\hz$ taking values
in the parameter space shown in Figure 1, and calculated the corresponding
values of $\tilde{B}$, which is defined by
\be
\label{Bt}
B = \frac{m^2\pi^2}{12\mu^2} \tilde{B} \,.
\ee
The results, with the thin-wall divergence, Eq.~(\ref{twb0}), factored out,
are plotted in Figure 2. We then fitted various analytic formulae to the
results, first for the special cases of the straight boundaries of the
parameter space and then for the full two-dimensional case.

\subsection{One-dimensional fitting}
\label{1df}
\subsubsection*{$\nu=0$ : a cubic potential}

We obtained $\tilde{B} = 27.62$ in the $\nu=\hz=0$
limit with the accuracy to the given figures,
and, from Eq.~(\ref{Bnz}), $\tilde{B} = 256 / 27$ in the
Hawking-Moss limit $(\hz = 1/2)$.
We compared the numerically calculated $\tilde{B}$ with
\be
\tilde{B}_{\rm fit} = 27.62 - 72.55 \hz^2 \,,
\ee
where the coefficients were chosen to give an exact fit at both
boundaries of the parameter region. In this case the relative error
$(\tilde{B}-\tilde{B}_{\rm fit})/\tilde{B}$ is found to be $< 0.6\%$.
A more accurate fit, again exact at both boundaries, is given by
\be
\tilde{B} = 27.62 - 71.02 \hz^2 - 24.48 \hz^6 \,.
\ee
Now the relative error becomes $<2.6\times 10^{-4}$.

\subsubsection*{$\epsilon=0$ : degenerate vacua}

The numerical results for this case have already been obtained
in Ref.~\cite{Basu} and applied to the nucleation of spherical
domain-wall bubbles during inflation.
{}From Eq.~(\ref{twe}) we can see that $\tilde{B} \rightarrow 4/\hz^3$
as $\hz \rightarrow 0$. Also, from Eq.~(\ref{Bno}), we get
$\tilde{B} = 64$ in the Hawking-Moss limit $(\hz = 1/\sqrt{8})$.
Thus the simplest fitting exact at both boundaries is
\be
\tilde{B} = \frac{4}{\hz^3}
 \l[ 1 - 8 \l(1-\frac{1}{\sqrt{2}}\r) \hz^2 \r] \,,
\ee
with a relative error $<2\%$.
A better fit, again exact at both boundaries, is given by
\be
\tilde{B} = \frac{4}{\hz^3}
 \l[ 1 - 1.887 \hz^2 - 3.649 \hz^4 \r] \,.
\ee
In this case, the relative error is $<6\times 10^{-4}$.

\subsubsection*{$\hz=0$ : flat space}

This case has already been examined in \cite{Adams}.
In our notation, his results are
\be
\tilde{B} = \frac{1}{\epsilon^3}
\l( 1 + a \epsilon + b \epsilon^2 \r) \,,
\ee
with
\be
a = 10.052 \;\;\;\;{\rm and}\;\;\;\; b = 16.612 \,.
\ee
This fitting formula is not insufficient but it has its largest
relative error $(\sim 0.17\%)$ at $\nu=0$.
However, the choice of parameters
\be
a = 10.07 \;\;\;\;{\rm and}\;\;\;\; b = 16.55 \,,
\ee
is better, making the fitting exact at both boundaries,
and giving a relative error $<0.14\%$.

\subsection{Two-dimensional fitting}

Now we try to give a fitting formula valid in the whole range of
parameter space.
Looking at the rather complicated form of Eq.~(\ref{twb1}),
we can guess that it is not easy to obtain a simple fitting formula.
Figures 3(a) and 3(b) are plots of $\tilde{B}_1$ and
$\tilde{B}_2$ which are defined by
\bea
\label{Bt1}
\tilde{B} & = & \frac{ 4 ( 1 + 10.07 \epsilon + 16.55 \epsilon^2 ) }
{ \delta \l( \epsilon + \delta \r)^{2} } \tilde{B}_1 \,, \\
\label{Bt2}
 & = & \frac{ 4 ( 1 + 10.07 \epsilon + 16.55 \epsilon^2 ) }
{ \delta \l( \epsilon + \delta \r)^{2} ( 1 + 10 \epsilon ) }
\l[ 1 + \frac{11}{2}\epsilon + 3\frac{\epsilon^2}{\delta}
+ \frac{3}{2}\frac{\epsilon^3}{\delta^2} \r] \tilde{B}_2 \,,
\eea
where $\delta \equiv \sqrt{\epsilon^2 + \hz^2}$.
{}From these plots, it seems better to try to fit $\tilde{B}_2$ instead
of $\tilde{B}$ or $\tilde{B}_1$, despite its more complicated factor.
Since this factor is written in terms of $\epsilon$ and $\delta$, we try
to fit $\tilde{B}_2$ with a polynomial of these two parameters.
To keep the relative error less than about 1\% , at least ten terms
seem to be necessary. We give one example of the fitting:
\be
\tilde{B}_2 = 1 + \sum_{ij} a_{ij} \epsilon^i \delta^j \,,
\ee
with nonzero coefficients
\bea
a_{01} &=& - 0.1617 \,,\;\;\; a_{03} = - 5.507 \,, \nn \\
a_{11} &=& -11.34   \,,\;\;\; a_{12} =  30.17  \,,\;\;\;
	a_{14} = -21.69 \,, \nn \\
a_{20} &=&  12.09   \,,\;\;\; a_{21} = -33.24  \,,\;\;\;
	a_{23} =  41.29 \,, \nn \\
a_{30} &=&   7.728  \,,\;\;\;{\rm and}\;\;\; a_{32} = -19.34  \,.
\eea
Factoring out the complicated thin-wall factor was necessary
to remove the divergence in the thin-wall limit.
Therefore, for small $\nu$, $\tilde{B}$ is expected to be fitted
by a simpler formula. We give the following example of a
fitting for $\nu \le 0.4$:
\be
\tilde{B} = \frac{1}{\epsilon^3}
\left( 1 + 10.07 \epsilon + 16.55 \epsilon^2 \right)
\left[ \frac{ 1 + (a+b\nu)\hz^2 }{ 1 + (c+d\nu+e\nu^2)\nu^2\hz^2 }
\right] \,,
\ee
with
\be
 a =  -  2.635 \,,\;\;\;
 b =  -  0.378 \,,\;\;\;
 c =    43.6   \,,\;\;\;
 d =  -130     \,,\;\;\;{\rm and}\;\;\;
 e =   163     \,.
\ee
The relative error of this fitting is $< 1.4\%$.

\section{Conclusions}

We have thoroughly investigated the tunneling rate for a scalar field
with a general quartic potential in de Sitter space in the limit of
no gravitational back reaction. In this case, the number of parameters
of the theory essentially reduces to two. We have numerically
calculated the tunneling rates over this two-dimensional parameter
space. Based on analytic formulae which are valid for extreme values of
the parameters, we have constructed approximate formulae from the
numerical results with relative errors of $\lsim 1\%$. Since these
approximate formulae incorporate the effect of the background
curvature, they have a wider range of applicability than previously
known formulae. Hence our results will provide a better understanding
of quantum tunneling phenomena and the associated physical processes
in the early Universe.

\section*{Acknowledgements}

EDS is supported by a JSPS Postdoctoral Fellowship and
TT is supported by a JSPS Junior Scientist Fellowship.
This work was supported by Monbusho Grant-in-Aid for Encouragement of
Young Scientists Nos.\ 2010 and 92062, Monbusho Grant-in-Aid for
Scientific Research No.\ 05640342, and the Sumitomo Foundation.

\section*{Figure Captions}

\subsection*{Figure 1: Parameter space}

The dotted line (given by Eq.~(\ref{hmbdy})) marks the boundary between
the regions where the
Hawking-Moss (HM) instanton (see Section~\ref{hm}) and our numerically
calculated instantons dominate. The corner labeled TW is where the
thin-wall approximation (see Section~\ref{tw}) is valid. The boundaries
labeled CP, FS, and DV correspond to the limits of a cubic potential,
flat space, and degenerate vacua, respectively (see Section~\ref{1df}).
The solid lines are a contour plot of our numerical results
(and of the analytic results in the Hawking-Moss region).
$\tilde{B}$ is defined in Eq.~(\ref{Bt}).

\subsection*{Figure 2: Numerical results}

The numerical results expressed in terms of $B/B_{\rm TW}$,
where $B_{\rm TW}$ is the thin-wall bounce action given in
Eq.~(\ref{twb0}).

\subsection*{Figure 3}

A comparison of two possible choices of functions to be fitted
by analytic formulae. $\tilde{B}_1$ and $\tilde{B}_2$ are defined
in Eqs.~(\ref{Bt1}) and (\ref{Bt2}), respectively.

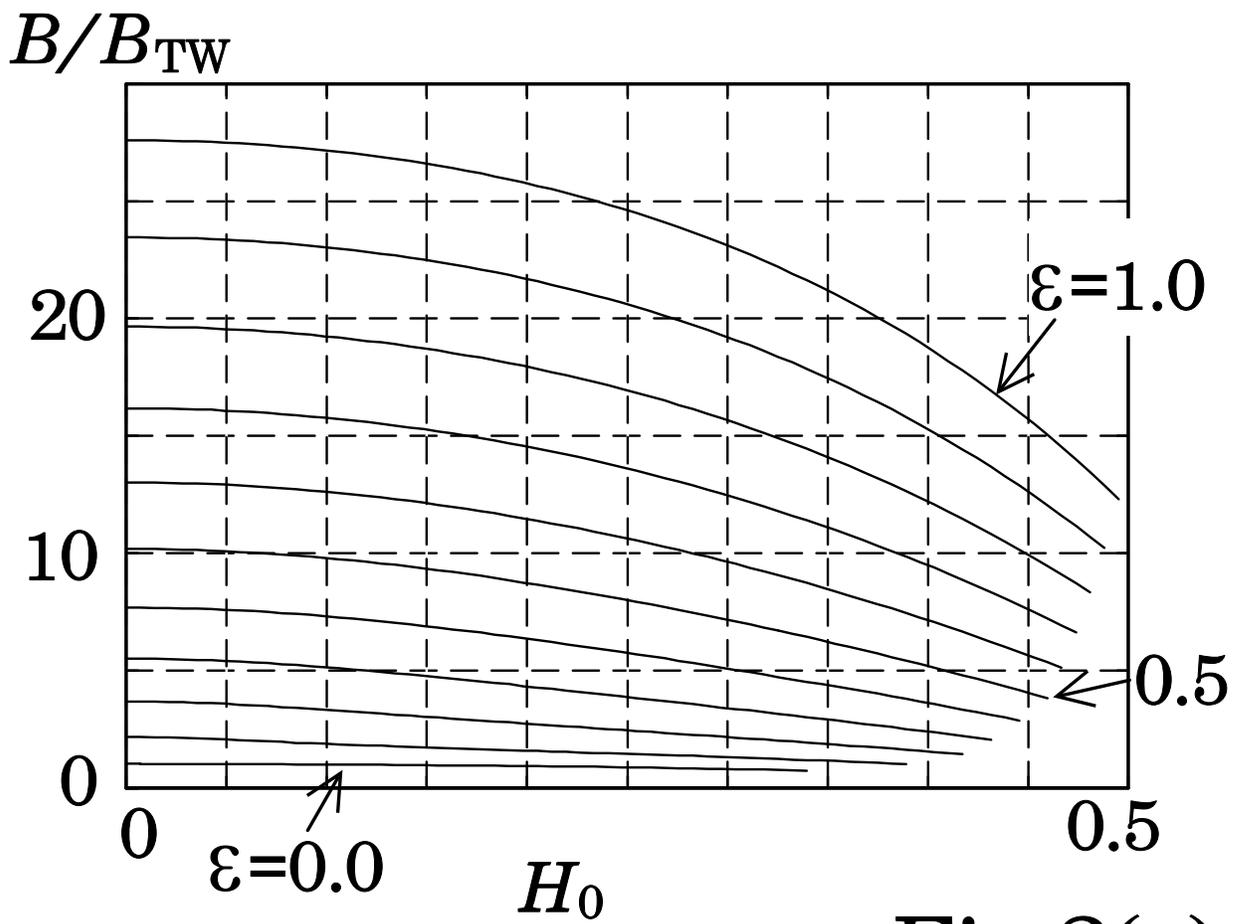
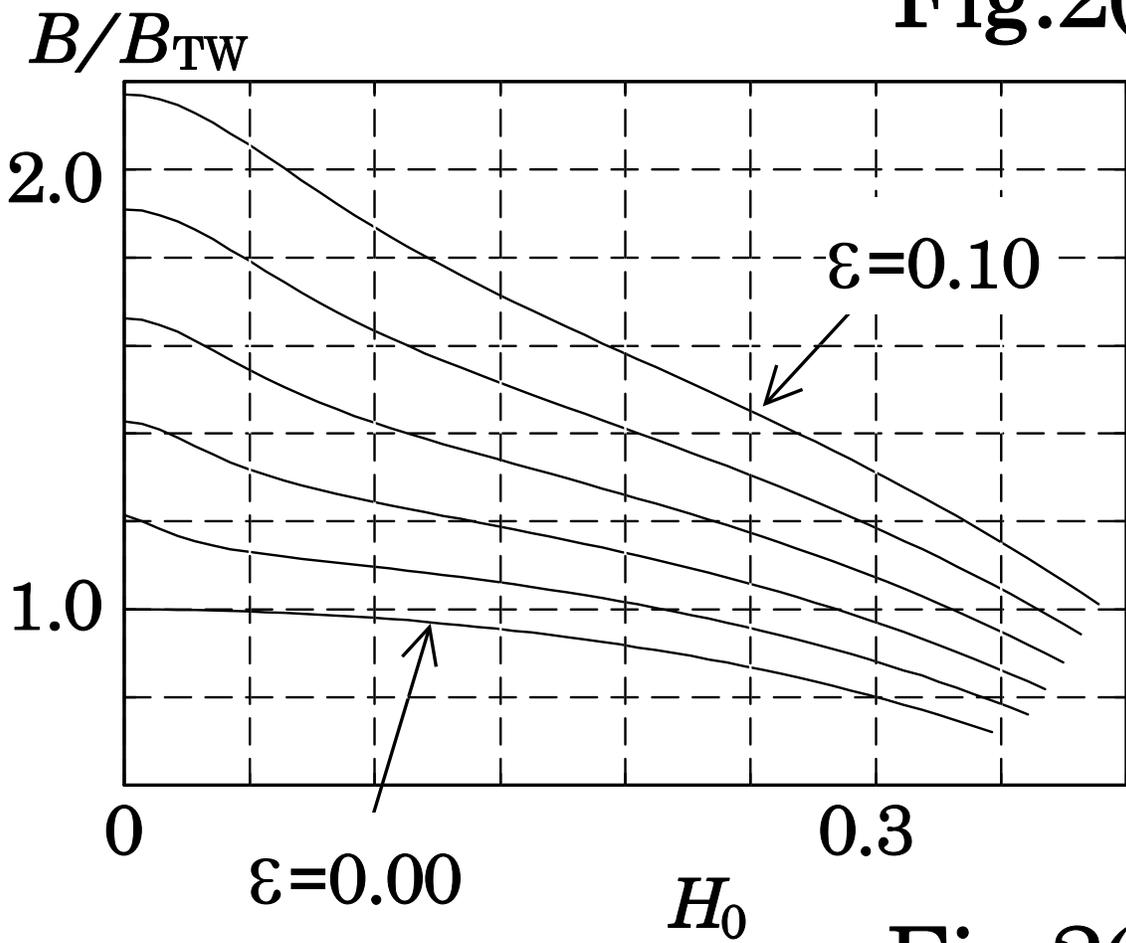
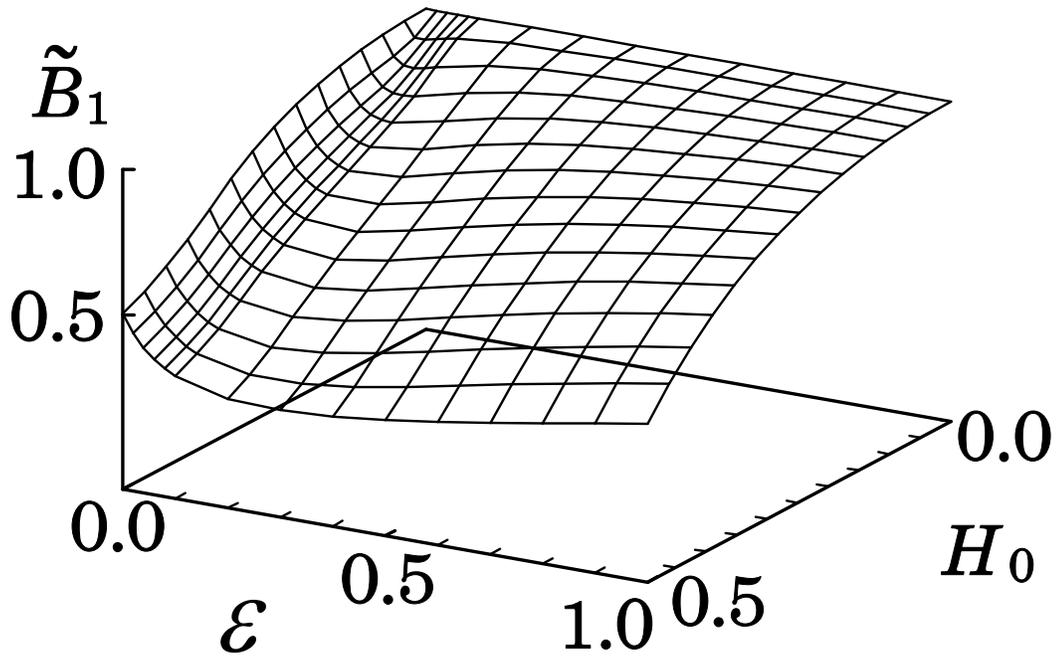
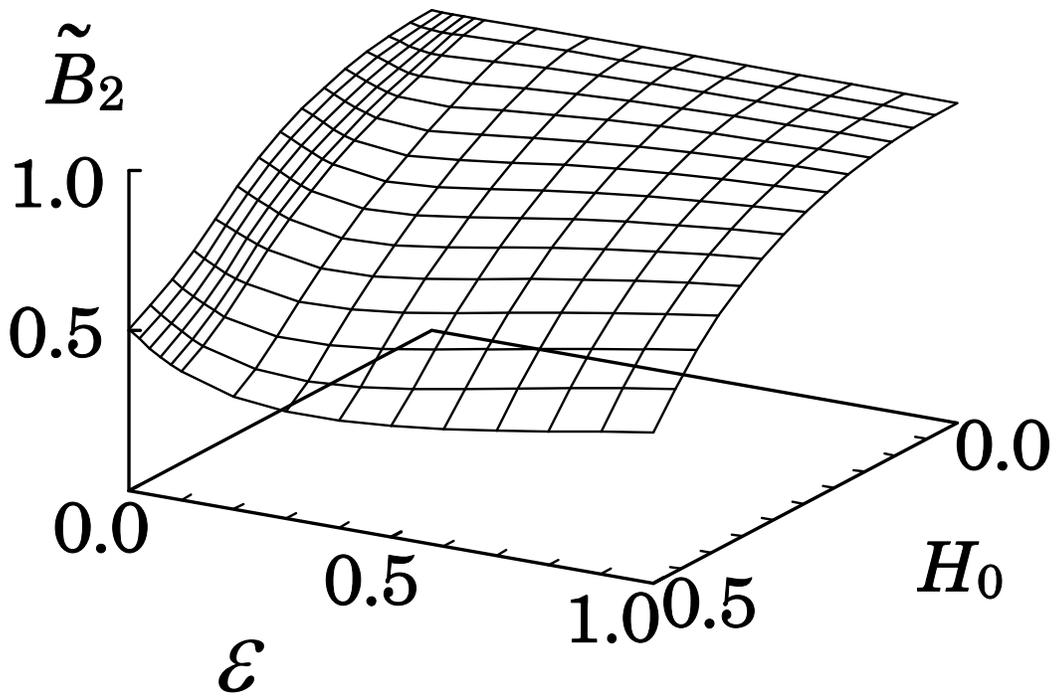
\end{document}